\documentclass[12pt, twoside]{article}
\usepackage[semicolon,round,sort&compress]{natbib}  %
\usepackage{hyperref}
\usepackage{amsmath,amsthm, amssymb, amsfonts}
\usepackage{booktabs}
\usepackage{pdflscape}
\usepackage{calc}
\usepackage{float}
\usepackage{framed}
\usepackage{indentfirst}
\usepackage{chronology}
\usepackage[toc,page]{appendix}
\usepackage{tablefootnote}
\usepackage{subfigure}
\graphicspath{{images/}}
\usepackage{caption}
\usepackage[a4paper,width=150mm,top=25mm,bottom=25mm,bindingoffset=6mm]{geometry} 
\usepackage{filecontents}
\usepackage{algorithm}
\usepackage{algorithmic}
\usepackage{nicefrac}
\usepackage{comment}

\usepackage{graphicx}
\usepackage{verbatim}
\usepackage{multirow}
\usepackage{bbm}
\usepackage[utf8]{inputenc}
\usepackage[english]{babel}
\usepackage{mathtools}
\usepackage{natbib}
\usepackage{mathtools}

\usepackage{multicol}
\renewcommand{\bibpreamble}{\begin{multicols}{2}}
\renewcommand{\bibpostamble}{\end{multicols}}

\makeatletter
\newcommand{\distas}[1]{\mathbin{\overset{#1}{\kern\z@\sim}}}%
\newsavebox{\mybox}\newsavebox{\mysim}
\newcommand{\distras}[1]{%
  \savebox{\mybox}{\hbox{\kern3pt$\scriptstyle#1$\kern3pt}}%
  \savebox{\mysim}{\hbox{$\sim$}}%
  \mathbin{\overset{#1}{\kern\z@\resizebox{\wd\mybox}{\ht\mysim}{$\sim$}}}%
}
\makeatother

\usepackage{changepage}
\definecolor{shadecolor1}{rgb}{0.40,0.83,0.70}
\definecolor{shadecolor2}{rgb}{0.84,0.83,0.63}

 {\endMakeFramed}

\newenvironment{shaded2}{%
  \MakeFramed {\FrameRestore}}%
 {\endMakeFramed}

\title{Rao--Blackwellization in the MCMC era}
\author{{\sc Christian P.~Robert$^{1,2,3}$ and Gareth O.~Roberts$^{1}$}\footnote{
The work of the first author was partly supported in
part by the French government under management of Agence Nationale de la Recherche as part of the ``Blanc SIMI 1'' program, reference ANR-18-CE40-0034 and in part by the French government under management of Agence Nationale de la Recherche as
part of the ``Investissements d’avenir'' program, reference ANR19-P3IA-0001 (PRAIRIE 3IA Institute).
The first author also acknowledges the support of l'Institut Universitaire de France through two consecutive senior chairs.}\\
{\em $^1$University of Warwick,}\\ 
{\em $^2$Université Paris Dauphine PSL},\\ 
{\em and $^3$CREST-ENSAE}}
\date{}

\begin{document}
\maketitle

\begin{abstract}
Rao--Blackwellization is a notion often occurring in the MCMC literature, with possibly different meanings and
connections with the original Rao--Blackwell theorem \citep{rao:1945,blackwell:1947}, including a reduction of
the variance of the resulting Monte Carlo approximations. This survey reviews some of the meanings of the term.
\end{abstract}


\begin{quote}\em
This paper is dedicated to Professor C.R. Rao in honour of his 100th birthday.
\end{quote}

\section{Introduction}

\begin{figure}[t]
\centerline{\includegraphics[width=.7\textwidth]{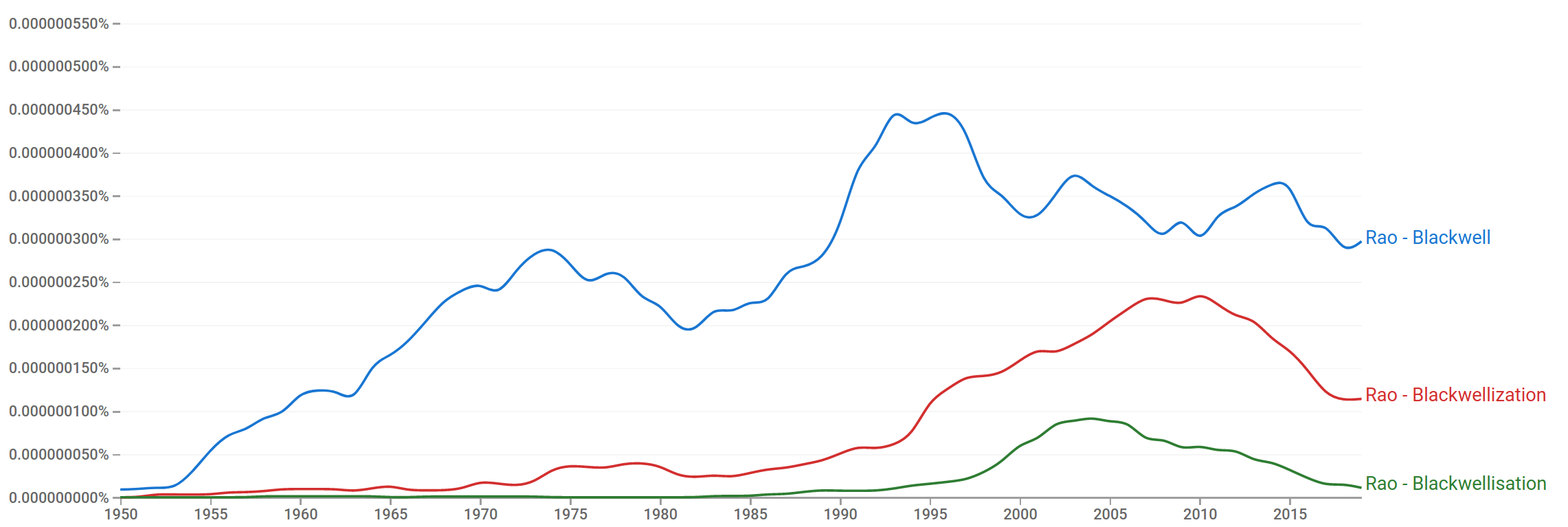}}
\caption{\label{fig:ngram}
Google assessment of the popularity of the names Rao--Blackwell, Rao--Blackwellisation, and Rao--Blackwellization since the derivation of the Rao--Blackwell theorem.}
\end{figure}

The neologism Rao--Blackwellization\footnote{We will use the American English
spelling of the neologism as this version is more commonly used in the
literature.}\footnote{\cite{berkson:1955} may have been the first one to use
(p.142) this neologism.} stems from the famous Rao--Blackwell theorem
\citep{rao:1945,blackwell:1947}, which states that replacing an estimator by
its conditional expectation given a sufficient statistic improves estimation
under any convex loss. This is a famous mathematical statistics result, both
dreaded and appreciated by our students for involving conditional expectation and
for producing a constructive improvement, respectively. While Monte Carlo
approximation techniques cannot really be classified as estimation, since they
operate over controlled simulations, rather than observations, with the ability
to increase the sample size if need be, and since there is rarely a free and
unknown parameter involved, hence almost never a corresponding notion of
sufficiency, seeking improvement in Monte Carlo approximation via partial
conditioning has nonetheless been named after this elegant theorem. As shown in
Figure \ref{fig:ngram}, the use of the expression Rao--Blackwellization has
considerably increased in the 1990's, once the foundational paper popularising
MCMC techniques refered to this technique to reduce Monte Carlo variability,

The concept indeed started in the \citeyear{gelfand:smith:1990} foundational
paper by Gelfand and Smith (``foundational" as it launched the MCMC revolution,
see \citealp{green:latuz:etal:2015}). While this is not exactly what is proposed in
the paper, as detailed in the following section, it is now perceived\footnote{See e.g. the comment
in the Introduction Section of \cite{liu:wong:kong:1994}.} that the
authors remarked that, given a Gibbs sampler whose component $\theta_1$ is simulated from the
conditional distribution, $\pi(\theta_1|\theta_2,x)$,
the estimation of the marginal $\pi(\theta_1|x)$ is improved by considering the average of
the (full) conditionals across iterations,
$$\nicefrac{1}{T} \sum_{t=1}^r \pi(\theta_1|\theta_2^{(t)},x)$$
which provides a parametric, unbiased and $\text{O}(\nicefrac{1}{\sqrt{T}})$
estimator. Similarly, the approximation to $\mathbb E[\theta_1|x]$ based on this representation
$$\nicefrac{1}{T} \sum_{t=1}^r \mathbb E[\theta_1|\theta_2^{(t)},x]$$
is using conditional expectations with lesser variances than the original
$\theta_1^{(t)}$ and may thus lead to a reduced variance for the estimator, if
correlation does not get into the way. (In that specific two-step sampler, this
is always the case \citealp{liu:wong:kong:1994}.

We are thus facing the difficult classification task of separating what is
Rao--Blackwellization from what is not Rao--Blackwellization in simulation and
in particular MCMC settings.

The difficulty resides in setting the limits as
\begin{itemize}
\item there is no clear notion of sufficiency in simulation and, further,
conditioning may increase the variance of the resulting estimator or slow down convergence;
\item variance reduction and unbiasedness are not always relevant (as shown by
the infamous harmonic mean estimator, \citealp{neal:1999,robert:wraith:2009}), as
for instance in infinite variance importance sampling \citep{chatterjee:diaconis:2018,vehtari:simpson:etal:2019};
\item there are (too) many forms of potential conditioning in simulation
settings to hope for a ranking (see, e.g., the
techniques of 
partitioning, antithetic or auxiliary variables, 
control variates as in \citealp{berg:zhu:clayton:2019}, 
delayed acceptance as in \citealp{banterle:etal:2019,beskos:papaspiliopoulos:roberts:fearnhead:2006},
adaptive mixtures as in \citealp{owen:zhou:00,
cornuet:marin:mira:robert:2012,elvira:etal:2019}, the later more closely connected to Rao--Blackwellization);
\item the large literature on the approximation of normalising constants and Bayes
factors \citep{robert:marin:2008,marin:robert:2010b,marin:robert:2010} contains many proposals that
relate to Rao-Blackwellisation, as, e.g., through the simulation of auxiliary samples from instrumental
distributions as initiated in \cite{geyer:1993} and expanded into bridge sampling by \cite{chopin:robert:2010}
and noise-constrastive estimation by \cite{gutmann:hyvarinen:2012};
\item in connection with the above, many versions of demarginalization such as
slice sampling \citep{roberts:rosenthal:1999,mira:moeller:roberts:2001} introduce 
auxiliary variables that could be exploited towards bringing a variance reduction;\footnote{This
may however be seen as a perversion of Rao--Blackwellization in that the dimension of the
random variable used in the simulation is increased, with the resulting estimate being obtained
by the so-called {\em Law of the Unconscious Statistician}.}
\item there is no optimal solution in simulation as, mathematically, a quantity
such as an expectation is uniquely and exactly defined once the distribution is
known: if computation time is not accounted for, the exact value is the optimal
solution;
\item while standing outside a probabilistic framework, quasi-Monte Carlo
techniques \citep{liao:1998} can also be deemed to constitute an ultimate form
of Rao--Blackwellization, with the proposal of \cite{kong:mccullagh:meng:2003}
being an intermediate solution;\footnote{As mentioned by the authors, the
``group-averaged estimator may be interpreted as Rao--Blackwellization given
the orbit, so group averaging cannot increase the variance" (p. 592)}
\end{itemize}
but we will not cover any further these aspects here.

The rest of this review paper discusses Gibbs sampling in Section \ref{sec:GibbS},
other MCMC settings in \ref{sec:MCMC}, particle filters and SMC in
\ref{sec:SMC}, and conclude in Section \ref{sec:onc}.

\section{Gibbs sampling}
\label{sec:GibbS}

\noindent 
Let us recall that a Gibbs sampler \citep{geman:geman:1984} is a specific way of
building a Markov chain with stationary density $\pi(\cdot)$ through the iterative generation 
from conditional densities associated with the joint $\pi(\cdot)$. Its simplest version consists in 
partitioning the argument $\theta$ into $\theta=(\theta_1,\theta_2)$ and generating alternatively
from $\pi_1(\theta_1|\theta_2)$ and from $\pi_2(\theta_2|\theta_1)$. This binary version is sometimes called
{\em data augmentation} in reference to \cite{tanner:wong:1987}, who implemented an algorithm related to
the Gibbs sampler for latent variable models.

When proposing this algorithm as a way to simulating from marginal densities and (hence) posterior distributions, 
Gelfand and Smith (1990) explicitely relate to the Rao--Blackwell theorem, as shown by the following quote\footnote{The
text has been retyped and may hence contains typos. The notations are those introduced by \cite{gelfand:smith:1990} and
used for a while in the literature, see e.g. \cite{spiegelhalter:thomas:best:gilks:1995} with $[X \mid Y]$ denoting the 
conditional density of $X$ given $Y$. The double indexation of the sequence is explained below.}

\begin{quotation}
\em ...we consider the problem of calculating a final form of marginal density from
the final sample produced by either the substitution or Gibbs sampling
algorithms. Since for any estimated marginal the corresponding full
conditional has been assumed available, efficient inference about the
marginal should clearly be based on using this full conditional distribution.
In the simplest case of two variables, this implies that $[X \mid Y]$ and the $y^{(i)}_j$'s
$(j = 1, \ldots, m)$ should be used to make inferences about $[X]$, rather than
imputing $X^{(i)}_j$ $(j = 1, \ldots, nm)$ and basing inference on these $X^{(i)}_j$'s. Intuitively,
this follows, because to estimate $[X]$ using the $x^{(i)}_j$'s requires a kernel density
estimate. Such an estimate ignores the known form $[X \mid Y]$ that is mixed to
obtain $[X]$. The formal argument is essentially based on the Rao--Blackwell
theorem. We sketch a proof in the context of the density estimator itself. If
$X$ is a continuous p-dimensional random variable, consider any kernel density
estimator of $[X]$ based on the $X^{(i)}_j$'s (e.g., see Devroye and Gy{\"o}rfi, 1985)
evaluated at $x_0$: $\Delta^{(i)}_{x_0} = (1/h_m^p) \sum_{j=1}^m K[(X_0 -
X^{(i)}_j)/h_m]$, say, where $K$ is a bounded density on $\mathbb R^p$ and the
sequence $\{h_m\}$ is such that as $m \rightarrow \infty$, $h_m \rightarrow 0$,
whereas $mh_m \rightarrow \infty$. To simplify notation, set $Q_{m,x_0}(X) =
(1/h_m^p) K[(X - X^{(i)}_j)/h_m]$
so that $\Delta^{(i)}_{x_0} = (1/m) \sum_{j=1}^m Q_{m,x_0}(X_j^{(i)})$.
Define $\gamma_{x_0}^i = (1/m) \sum_{j=1}^m \mathbb E[Q_{m,x_0}(X)\mid Y^{(i)}_j]$. By
our earlier theory, both $\Delta^{(i)}_{x_0}$ and $\gamma_{x_0}^i$ have the
same expectation. By the Rao--Blackwell theorem, $\text{var}\, \mathbb E[Q_{m,x_0}(X)]
\mid Y) \le \text{var}\, Q_{m,x_0}(X)$, and hence $\text{MSE}(\gamma_{x_0}^i)\le
\text{MSE}(\Delta^{(i)}_{x_0})$, where MSE denotes the mean squared error of the estimate of 
$[X]$.
\end{quotation}

This Section 2.6. of the paper calls for several precisions:
\begin{itemize}
\item the simulations $x^{(i)}_j$ and $y^{(i)}_j$ are double-indexed because the authors
consider $m$ parallel and independent runs of the Gibbs sampler, $i$ being the number of iterations
since the initial step, in continuation of \cite{tanner:wong:1987},
\item the Rao--Blackwell argument is more specifically a conditional expectation step,
\item as later noted by \cite{geyer:1995}, the conditioning argument is directed at (better) 
approximating the entire density $[X]$, even though the authors mention on the
following page that the argument is ``simpler for estimation of" a posterior expectation,
\item they compare the mean squared errors of the expected density estimate rather
than the rates of convergence of a non-parametric kernel estimator(in
$n^{\nicefrac{-1}{4+d}}$) versus an unbiased parametric density estimator
(in $n^{\nicefrac{-1}{2}}$), which does not call for a Rao--Blackwellization argument,
\item they do not (yet) mention ``Rao--Blackwellization" as a technique,
\item and they do not envision (more) ergodic averages across iterations,
possibly fearing the potential impact of the correlation between the terms
for a given chain.
\end{itemize}

A more relevant step in the use of Rao--Blackwellization techniques for the Gibbs sampler
is found in \cite{liu:wong:kong:1994}. This later article establishes in particular that, for the two-step Gibbs 
sampler, Rao--Blackwellization always produces a decrease in the variance of the empirical averages. This
is established in a most elegant manner by showing that each extra conditioning
(or further lag) decreases the correlation, which is always
positive.\footnote{The ``always" qualification applies to every transform of
the chain and to every time lag.} The proof relies on the associated notion of
{\em interleaving} and expresses the above correlation as the variance of a
multiply conditioned expectation: 
$$
\text{cov}(h(\theta_1^{(0)}), h(\theta_1^{(n)}) = \text{var}\,(\mathbb E(\ldots
\mathbb E[ \mathbb E\{ h(\theta_1) \mid \theta_2 \}] \ldots )\}\,,
$$ 
where the number of conditional expectations on the rhs is $n$. The authors also warn that a ``fast
mixing scheme gains an extra factor in efficiency if the mixture estimate can
be easily computed" and give a counter-example when Rao--Blackwellization
increases the variance.\footnote{The function leading to the counter-example is
however a function of both $\theta_1$ and $\theta_2$, which may be considered
as less relevant in latent variable settings.} This counter-example is
exploited in a contemporary paper by \cite{geyer:1995}\footnote{\cite{geyer:1995}
also points out that a similar Rao--Blackwellization was proposed by
\cite{pearl:1987}.} where a necessary and sufficient but highly theoretical
condition is given for an improvement. As the author puts it in his conclusion,
\begin{quotation}
\em The point of this article is not that Rao--Blackwellized estimators are a good thing or a
bad thing. They may be better or worse than simple averaging of the functional
of interest without conditioning. The point is that, when the autocorrelation
structure of the Markov chain is taken into account, it is not a theorem that
Rao--Blackwellized estimators are always better than simple averaging. Hence the
name Rao--Blackwellized should be avoided, because it brings to mind optimality
properties that these estimators do not really possess. Perhaps ``averaging a
conditional expectation" is a better name.
\end{quotation}
but his recommendation was not particularly popular, to judge from the subsequent literature 
resorting to this denomination.

Another connection between Rao--Blackwellization and Gibbs sampling can be found in \cite{chib:1995}, where his 
approximation to the marginal likelihood
$$m(x) = \dfrac{\pi(\theta^*)f(x|\theta^*)}{\hat\pi(\theta^*|x)}$$
is generaly based on an estimate\footnote{\cite{chib:1995} mentions this connection (p.1314) 
but seems to restrict it to the two-stage Gibbs sampler. In the earlier version known as ``the candidate's
formula", due to a Durham student coming up with it, \cite{besag:candidate:1989} points out the possibility
of using an approximation such as a Laplace approximation, rather than an MCMC estimation.}\footnote{A question
found on the statistics forum Cross-Validated illustrates the difficulty with understanding demarginalisation
and joint simulation: {\em ``Chib suggests that we can insert the Gibbs sampling outputs of $\mu$ into the
summation {\em [of the full conditionals]}. But aren't the outputs obtained from Gibbs about the joint posterior
$p(\mu,\phi|y)$? Why suddenly can we use the results from joint distribution to
replace the marginal distribution?"}}
of the posterior density using a latent (or auxiliary)
variable, as in \cite{gelfand:smith:1990},
$$\hat\pi(\theta^*|x)=\nicefrac{1}{T} \sum_{t=1}^T \pi(\theta^*|x,z^{(t)})$$
The stabilisation brought by this parametric approximation is notable when compared with kernel estimates, 
even though it requires that the marginal distribution on $z$ is correctly simulated \citep{neal:1999}.

\section{Markov chain Monte Carlo methods}
\label{sec:MCMC}

\noindent 
In the more general setting of Markov chain Monte Carlo (MCMC) algorithms
\citep{robert:casella:2004}, further results characterise
the improvement brought by Rao--Blackwellization. Let us briefly recall that the concept behind
MCMC is 
to create a Markov sequence $\theta^{(n)})_n$ of dependent variables that
converge (in distribution) to the distribution of interest (also called {\em target}). One of the most
ubiquitous versions of an MCMC algorithm is the Metropolis--Hastings algorithm 
\citep{metropolis:1953,hastings:1970,green:latuz:etal:2015}

One direct exploitation of the Rao--Blackwell theorem is found in \cite{mckeague:wefelmeyer:2000},
who sho in particular that, when estimating the mean of $\theta$ under the target distribution,
 a Rao--Blackwellized version based on $\mathbb E[h(\theta^{(n)}|\theta^{(n-1)}]$ will improve the
asymptotic variance of the ordinary empirical estimator when the chain $(\theta^{(n)})$ 
is reversible. While the setting may appear quite restrictive, the authors manage to recover data augmentation
with a double conditional expectation (when compared with \citealp{liu:wong:kong:1994}) as well as reversible
Gibbs and Metropolis samplers of the Ising model. The difficulty in applying the method resides in computing
the conditional expectation, since a replacement with a Monte Carlo approximation cancels its appeal.

\cite{casella:robert96} consider an altogether different form of
Rao--Blackwellization for both accept-reject and Metropolis--Hastings samples.
The core idea is to integrate out via a global conditional expectation the Uniform
variates used to accept or reject the proposed values.

A sample produced by the Metropolis--Hastings algorithm,
$\theta^{(1)},\ldots,\allowbreak \theta^{(T)}$, is in fact based on two simulated samples,
the sample of proposed values $\eta_1,\ldots,\eta_T$ and the sample of decision variates
$u_1,\ldots,u_T$, with $\eta_t\sim q(y|\theta^{(t-1)})$ and
$u_t \sim \mathcal U([0,1])$. Since $\theta^{(t)}$ is equal to one of the earlier proposed values,
an empirical average associated with this sample can be written\footnote{In order to avoid additional
notations, we assume a continuous model where all $\eta_i$'s are different with probability one.}
$$
\delta^{MH} = \nicefrac{1}{T} \sum_{t=1}^T h(\theta^{(t)}) 
= \nicefrac{1}{T} \sum_{t=1}^T \sum_{i=1}^t \; \mathbb I_{\theta^{(t)}=\eta_i} \; h(\eta_i)\,.
$$
Therefore, taking a conditional expectation of the above by integrating the decision variates,
\begin{align*}
\delta^{RB} &= \displaystyle{\nicefrac{1}{T}\sum_{i=1}^T \; h(\eta_i) \;
\mathbb E\left[\sum_{t=i}^T \; \mathbb I_{\theta^{(t)}=\eta_i}\bigg|\eta_1,\ldots,\eta_T\right] } \\
&= \displaystyle{\nicefrac{1}{T} \sum_{i=1}^T \; h(\eta_i) \; 
\sum_{t=i}^T \; \mathbb P(\theta^{(t)}=\eta_i|\eta_1,\ldots,\eta_T)}\,,
\end{align*}
leads to an improvement of the empirical average, $\delta^{MH}$, under convex losses.

While, for the independent Metropolis--Hastings algorithm, the conditional probability
can be obtained in closed form (see also \citealp{atchade:perron:2005} and
\citealp{jacob:robert:smith:2010}), the general case, based on an arbitrary
proposal distribution $q(\cdot|\theta)$ is such that $\delta^{RB}$ is less tractable but
\cite{casella:robert96} derive a tractable recursive expression for the weights 
of $h(\eta_i)$ in $\delta^{RB}$, with complexity of order $\mathcal{O}(T^2)$. Follow-up papers
are \cite{perron:1999} and \cite{casella:robert:wells:2004b}.

While again attempting at integrating out the extraneous uniform variates exploited 
by the Metropolis--Hastings algorithm, \cite{douc:robert:2010} derive another Rao--Blackwellized 
improvement over the regular Metropolis--Hastings algorithm by following a different representation
of $\delta^{MH}$, using the accepted chain $(\xi_i)_i$ instead of the proposed
sequence of the $\eta_i$'s as in \cite{casella:robert96}. The version based on accepted values is
indeed rewritten as
$$
\delta^{MH} = \nicefrac{1}{T}\,\sum_{i=1}^M \mathfrak{n}_i h(\xi_i)\,,
$$
where the $\xi_i$'s are the accepted $\eta_j$'s, $M$ is the number of accepted $\eta_j$'s till iteration $T$, and
$\mathfrak{n}_i$ is the number of times $\xi_i$ appears in the sequence $(\theta^{(t)})_t$.
This representation is also exploited in \cite{sahu:zhigljavsky:1998,
gasemyr:2002,sahu:zhigljavsky:2003}, and \cite{malefaki:iliopoulos:2008}. The Rao--Blackwellisation
construct of \cite{douc:robert:2010} exploits the following properties:
\begin{enumerate}
\item $(\xi_i,\mathfrak n_i)_i$ is a Markov chain;
\item $\xi_{i+1}$ and $\mathfrak n_i$ are independent given $\xi_i$;
\item $\mathfrak n_i$ is distributed as a Geometric random variable with probability parameter
\begin{equation}\label{eq:defp}
p(\xi_i) := \int \alpha(\xi_i,\eta)\,q(\eta|\xi_i)\,\mathrm{d}\eta\,;
\end{equation}
\item $(\xi_i)_i$ is a Markov chain with transition kernel $\tilde
Q(\xi,\mathrm{d}\eta)=\tilde q(\eta|\xi)\mathrm{d}\eta$ and stationary
distribution $\tilde \pi$ such that 
$$
\tilde q(\cdot|\xi) \propto  \alpha(\xi,\cdot)\,q(\cdot|\xi)\quad \mbox{and}
\quad  \tilde \pi(\cdot) \propto \pi(\cdot)p(\cdot)\,.
$$
\end{enumerate}

Since the Metropolis--Hastings estimator $\delta^{MH}$ only involves the $\xi_i$'s, i.e.~the accepted $\eta_t$'s,
an optimal weight for those random variables is the importance weight $1/p(\xi_i)$,
leading to the corresponding importance sampling estimator
$$
\delta^{IS} = \nicefrac{1}{N}\,\sum_{i=1}^M \dfrac{h(\xi_i)}{p(\xi_i)}\,,
$$
but this quantity is almost invariably unavailable in closed form and need be
estimated by an unbiased estimator. The geometric $\mathfrak n_i$ is the {\em de facto} solution that is used
in the original Metropolis-Hastings estimate, but solutions with smaller variance also are available, based on the
property that (if $\alpha(\xi,\eta)$ denotes the Metropolis--Hastings acceptance probability)
$$
\hat\zeta_i = 1+\sum_{j=1}^\infty \prod_{\ell\le j} \left\{ 1 - \alpha(\xi_i,\eta_\ell) \right\}
\qquad \eta_\ell\stackrel{\text{i.i.d.}}{\sim}q(\eta|\xi_i)
$$
is an unbiased estimator of $1/p(\xi_i)$ whose variance, conditional on $\xi_i$,
is lower than the conditional variance of $\mathfrak n_i$, $\{1-p(\xi_i)\}/p^2(\xi_i)$.
For practical implementation, in the event $\alpha(\xi,\eta)$ is too rately equal to one,
the number of terms where the indicator funtion is replaced with
its expectation $\alpha(\xi_i,\eta_\ell)$ may be limited, without jeopardising the variance
domination.

\section{Retrospective: Continuous time Monte Carlo methods}
\label{sec:CTMC}

Retrospective simulation \citep{beskos:papaspiliopoulos:roberts:fearnhead:2006}
is an attempt to take advantage of the redundancy inherent in modern simulation algorithms 
(particularly MCMC, rejection sampling) by subverting the traditional order of algorithm steps.
It is connected to demarginalisation and pseudo-marginal \citep{andrieu:roberts:2009} techniques
in that it replaces a probability of acceptance with an unbiased estimation of the said probability,
hence creating an auxiliary variable in the process. In the case of the Metropolis-Hastings algorithm, 
this means substituting the ratio
$$\dfrac{\pi(\theta^\prime) q(\theta^{(t)}|\theta^\prime)}{\pi(\theta^{(t)}) q(\theta^\prime|^{(t)})}$$
with
$$\dfrac{\hat\pi^\prime q(\theta^{(t)}|\theta^\prime)}{\hat\pi^{(t)} q(\theta^\prime|^{(t)})}$$
where $\hat\pi^\prime$ is an auxiliary variable such that 
$$\mathbb E[\hat\pi^\prime|\theta^\prime]=\kappa\pi(\theta^\prime)
\qquad
\mathbb E[\hat\pi^{(t)}|\theta^{(t)}]=\kappa\pi(\theta^{(t)})$$
Retrospective simulation is most powerful in infinite dimensional contexts, where its natural
competitors are approximate and computationally expensive. The solution advanced by
\cite{beskos:papaspiliopoulos:roberts:fearnhead:2006} and \cite{beskos:papa:roberts:2006} to simulate
diffusions in an exact manner (for a finite number of points) relies on an auxiliary and bounding
Poisson process. The selected points produced this way actually act as a random sufficient statistic
in the sense that the stochastic process can be generated from Brownian bridges between these points
and closed form estimators conditional of these points may be available and with a smaller variance. See
also \cite{fearnhead:etal:2017} for related results on continuous-time
importance sampling (CIS). This includes a sequential importance sampling
procedure with a random variable whose expectation is equal to the importance
weight.\footnote{One difficulty with the approach is the possible occurrence of negative importance weights
\citep{jacob:thiery:2015}.}

\section{Rao--Blackwellized particle filters}
\label{sec:SMC}

Also known as particle filtering\footnote{An early instance, called bootstrap
filter \citep{gordon:salmon:smith:1993}, involved one of the authors of
\cite{gelfand:smith:1990}, who thus contributed to the birth of two major
advances in the field.} sequential Monte Carlo \citep{liu:chen:1998,doucet99,delmoral:doucet:jasra:2006} 
is another branch of the Monte Carlo methodology where the concept of Rao--Blackwellisation
has had an impact. We briefly recall here that sequential Monte Carlo is used in state-space and
other Bayesian dynamic models where the magnitude of the latent variable prevents the call to traditional
Monte Carlo (and MCMC) techniques. It is also relevant for dealing with complex static problems by creating
a sequence of intermediate and artificial models, a technique called tempering \citep{marinarietparisi92}.

\cite{doucet:defreitas:murphy:2000} introduce a general version of the Rao--Blackwellized particle filter by
commenting on the inherent inefficiency of particle filters in large dimensions, compounded by the dynamic
nature of the sampling scheme. The central filtering equation is a Bayesian update of the form
\begin{equation}\label{eq:pf}
p(z_{1:t}|y_{1:t}\propto p(z_{1:(t-1)}|y_{1:(t-1)}) p(y_t|z_t)p(z_t|z_{t-1})
\end{equation}
in a state-space formulation where $(z_t)_t$ (also denoted $z_{1:T}$) is the
latent Markov chain and $(y_t)_t$ the observed sequence. In this update, the conditional densities of $z_{1:t}$
and $z_{1:(t-1)}$ are usually unavailable and need be approximated by sampling solutions.

If some marginalisation of the sampling is available for the model at hand, this
reduces the {\em degeneracy phenomenon} at the core of particle filters. The example
provided in \cite{doucet:defreitas:murphy:2000} is one where $z_t(x_t,r_t)$, with 
$$p(x_{1:t},r_{1:t}|y_{1:t})=p(x_{1:t}|y_{1:t},r_{1:t})p(r_{1:t}|y_{1:t})$$
and $p(x_{1:t}|y_{1:t},r_{1:t})$ available in closed form.\footnote{\cite{doucet:defreitas:murphy:2000}
provide a realistic illustration for a neural network where the manageable part is obtained via a Kalman filter.} 
This component can then be used in the approximation of the filtering
distribution \eqref{eq:pf}, instead of weighted Dirac masses, which improves
its precision if only by bringing a considerable decrease in the dimension of
the particles \citep[Proposition 2]{doucet:defreitas:murphy:2000}.  It is
indeed sufficient to resort only to particles for the intractable part.

See \cite{andrieu:defreitas:doucet:2001,johansen:whiteley:doucet:2012,
Lindsten:2011,lindsten:schon:olsson:2011} for further extensions on this
principle. In particular, the PhD thesis of \cite{Lindsten:2011} contains the 
following and relevant paragraph:

\begin{quotation}\em
Moving from {\em [the particle estimator]} to {\em [its Rao--Blackwellized
version]} resembles a Rao--Blackwellisation of the estimator (see also Lehmann,
1983). In some sense, we movefrom a Monte Carlo integration to a partially
analytical integration. However, it is not clear that the Rao--Blackwellized
particle filter truly is a Rao-Blackwellisation of {\em [the original]}, in the
factual meaning of the concept.  That is, it is not obvious that the
conditional expectation of {\em [the original]} results in the {\em [its
Rao--Blackwellized version]}.  This is due to the nontrivial relationship
between the normalised weights generated by the {\em [particle filter]}, and
those generated by {\em [its Rao--Blackwellized version]}.  It can thus be said
that {\em [it]} has earned its name from being inspired by the Rao--Blackwell
theorem, and not because it is a direct application of it.
\end{quotation}

Nonetheless, any exploitation of conditional properties that does not induce a (significant) bias 
is bound to bring stability and faster convergence to particle filters.

\section{Conclusion}
\label{sec:onc}

The term of Rao--Blackwellisation is therefore common enough in the MCMC
literature to be considered as a component of the MCMC toolbox. As we pointed
out in the introduction, many tricks and devices introduced in the past can
fall under the hat of that term and, while a large fraction of them does not
come with a demonstrated improvement over earlier proposals, promoting the
concepts of conditioning and demarginalising as central to the field should be
seen as essential for researchers and students alike. Linking such concepts,
shared by statistics and Monte Carlo, with an elegant and historical result
like the Rao--Blackwell theorem stresses both the universality and the
resilience of the idea.

\small
\bibliographystyle{apalike}
\hyphenation{Post-Script Sprin-ger}


\begin{thebibliography}{}

\bibitem[Andrieu et~al., 2001]{andrieu:defreitas:doucet:2001}
Andrieu, C., de~Freitas, N., and Doucet, A. (2001).
\newblock Rao--{B}lackwellised particle filtering via data augmentation.
\newblock In {\em Advances in Neural Information Processing Systems (NIPS)},
  pages 561--567.

\bibitem[Andrieu and Roberts, 2009]{andrieu:roberts:2009}
Andrieu, C. and Roberts, G. (2009).
\newblock The pseudo-marginal approach for efficient {M}onte {C}arlo
  computations.
\newblock {\em Ann. Statist.}, 37(2):697--725.

\bibitem[Atchad\'e and Perron, 2005]{atchade:perron:2005}
Atchad\'e, Y. and Perron, F. (2005).
\newblock Improving on the independent {M}etropolis--{H}astings algorithm.
\newblock {\em Statistica Sinica}, 15:3--18.

\bibitem[Banterle et~al., 2019]{banterle:etal:2019}
Banterle, M., Grazian, C., Lee, A., and Robert, C.~P. (2019).
\newblock Accelerating {M}etropolis--{H}astings algorithms by delayed
  acceptance.
\newblock {\em Foundations of Data Science}, 1(2):103.

\bibitem[Berg et~al., 2019]{berg:zhu:clayton:2019}
Berg, S., Zhu, J., and Clayton, M.~K. (2019).
\newblock Control variates and {R}ao--{B}lackwellization for deterministic
  sweep {M}arkov chains.
\newblock arXiv:1912.06926.

\bibitem[Berkson, 1955]{berkson:1955}
Berkson, J. (1955).
\newblock Maximum likelihood and minimum $\chi^2$ estimates of the logistic
  function.
\newblock {\em J. American Statist. Assoc.}, 50(269):130--162.

\bibitem[Besag, 1989]{besag:candidate:1989}
Besag, J. (1989).
\newblock {A candidate's formula: A curious result in {B}ayesian prediction}.
\newblock {\em Biometrika}, 76(1):183--183.

\bibitem[Beskos et~al., 2006a]{beskos:papaspiliopoulos:roberts:fearnhead:2006}
Beskos, A., Papaspiliopoulos, O., Roberts, G., and Fearnhead, P. (2006a).
\newblock Exact and computationally efficient likelihood-based estimation for
  discretely observed diffusion processes (with discussion).
\newblock {\em J. Royal Statist. Society Series B}, 68(3):333--382.

\bibitem[Beskos et~al., 2006b]{beskos:papa:roberts:2006}
Beskos, A., Papaspiliopoulos, O., and Roberts, G.~O. (2006b).
\newblock Retrospective exact simulation of diffusion sample paths with
  applications.
\newblock {\em Bernoulli}, 12(6):1077--1098.

\bibitem[Blackwell, 1947]{blackwell:1947}
Blackwell, D. (1947).
\newblock Conditional expectation and unbiased sequential estimation.
\newblock {\em Ann. Statist.}, 18(1):105--110.

\bibitem[Casella and Robert, 1996]{casella:robert96}
Casella, G. and Robert, C. (1996).
\newblock Rao-{B}lackwellization of sampling schemes.
\newblock {\em Biometrika}, 83:81--94.

\bibitem[Casella et~al., 2004]{casella:robert:wells:2004b}
Casella, G., Robert, C.~P., and Wells, M.~T. (2004).
\newblock Generalized accept-reject sampling schemes.
\newblock {\em Lecture Notes-Monograph Series}, 45:342--347.

\bibitem[Chatterjee and Diaconis, 2018]{chatterjee:diaconis:2018}
Chatterjee, S. and Diaconis, P. (2018).
\newblock The sample size required in importance sampling.
\newblock {\em Ann. Appl. Probab.}, 28(2):1099--1135.

\bibitem[Chib, 1995]{chib:1995}
Chib, S. (1995).
\newblock Marginal likelihood from the {G}ibbs output.
\newblock {\em J. American Statist. Assoc.}, 90:1313--1321.

\bibitem[Chopin and Robert, 2010]{chopin:robert:2010}
Chopin, N. and Robert, C. (2010).
\newblock Properties of nested sampling.
\newblock {\em Biometrika}, 97:741--755.

\bibitem[Cornuet et~al., 2012]{cornuet:marin:mira:robert:2012}
Cornuet, J.-M., Marin, J.-M., Mira, A., and Robert, C. (2012).
\newblock Adaptive multiple importance sampling.
\newblock {\em Scandinavian Journal of Statistics}, 39(4):798--812.

\bibitem[Del~Moral et~al., 2006]{delmoral:doucet:jasra:2006}
Del~Moral, P., Doucet, A., and Jasra, A. (2006).
\newblock Sequential {M}onte {C}arlo samplers.
\newblock {\em J. Royal Statist. Society Series B}, 68(3):411--436.

\bibitem[Douc and Robert, 2011]{douc:robert:2010}
Douc, R. and Robert, C. (2011).
\newblock A vanilla variance importance sampling via population {M}onte
  {C}arlo.
\newblock {\em Ann. Statist.}, 39(1):261--277.

\bibitem[Doucet et~al., 1999]{doucet99}
Doucet, A., de~Freitas, N., and Gordon, N. (1999).
\newblock {\em Sequential {MCMC} in Practice}.
\newblock Springer-Verlag.

\bibitem[Doucet et~al., 2000]{doucet:defreitas:murphy:2000}
Doucet, A., Freitas, N.~d., Murphy, K.~P., and Russell, S.~J. (2000).
\newblock Rao--{B}lackwellised particle filtering for dynamic {B}ayesian
  networks.
\newblock In {\em Proceedings of the 16th Conference on Uncertainty in
  Artificial Intelligence}, UAI '00, pages 176--183, San Francisco, CA, USA.
  Morgan Kaufmann Publishers Inc.

\bibitem[Elvira et~al., 2019]{elvira:etal:2019}
Elvira, V., Martino, L., Luengo, D., and Bugallo, M. (2019).
\newblock Generalized multiple importance sampling.
\newblock {\em Statist. Science}, 34(1):129--155.

\bibitem[Fearnhead et~al., 2017]{fearnhead:etal:2017}
Fearnhead, P., {\L}atuszynski, K., Roberts, G.~O., and Sermaidis, G. (2017).
\newblock Continious-time importance sampling: {M}onte {C}arlo methods which
  avoid time-discretisation error.
\newblock arXiv:1712.06201.

\bibitem[G\.{a}semyr, 2002]{gasemyr:2002}
G\.{a}semyr, J. (2002).
\newblock {M}arkov chain {M}onte {C}arlo algorithms with independent proposal
  distribution and their relation to importance sampling and rejection
  sampling.
\newblock Technical Report~2, Department of Statistics, Univ. of Oslo.

\bibitem[Gelfand and Smith, 1990]{gelfand:smith:1990}
Gelfand, A. and Smith, A. (1990).
\newblock Sampling based approaches to calculating marginal densities.
\newblock {\em J. American Statist. Assoc.}, 85:398--409.

\bibitem[Geman and Geman, 1984]{geman:geman:1984}
Geman, S. and Geman, D. (1984).
\newblock Stochastic relaxation, {G}ibbs distributions and the {B}ayesian
  restoration of images.
\newblock {\em IEEE Trans. Pattern Anal. Mach. Intell.}, 6:721--741.

\bibitem[Geyer, 1993]{geyer:1993}
Geyer, C. (1993).
\newblock Estimating normalizing constants and reweighting mixtures in {M}arkov
  chain {M}onte {C}arlo.
\newblock Technical Report 568, School of Statistics, Univ. of Minnesota.

\bibitem[Geyer, 1994]{geyer:1995}
Geyer, C. (1994).
\newblock Conditioning in {M}arkov chain {M}onte {C}arlo.
\newblock {\em J. Comput. Graph. Statis.}, 4:148--154.

\bibitem[Gordon et~al., 1993]{gordon:salmon:smith:1993}
Gordon, N., Salmond, J., and Smith, A. (1993).
\newblock A novel approach to non-linear/non-{G}aussian {B}ayesian state
  estimation.
\newblock {\em IEEE Proceedings on Radar and Signal Processing}, 140:107--113.

\bibitem[Green et~al., 2015]{green:latuz:etal:2015}
Green, P.~J., {\L}atuszy{\'n}ski, K., Pereyra, M., and Robert, C.~P. (2015).
\newblock Bayesian computation: a summary of the current state, and samples
  backwards and forwards.
\newblock {\em Statistics and Computing}, 25(4):835--862.

\bibitem[Gutmann and Hyv\"{a}rinen, 2012]{gutmann:hyvarinen:2012}
Gutmann, M.~U. and Hyv\"{a}rinen, A. (2012).
\newblock Noise-contrastive estimation of unnormalized statistical models, with
  applications to natural image statistics.
\newblock {\em J. Mach. Learn. Res.}, 13(1):307--361.

\bibitem[Hastings, 1970]{hastings:1970}
Hastings, W. (1970).
\newblock {M}onte {C}arlo sampling methods using {M}arkov chains and their
  application.
\newblock {\em Biometrika}, 57:97--109.

\bibitem[Jacob et~al., 2011]{jacob:robert:smith:2010}
Jacob, P., Robert, C., and Smith, M. (2011).
\newblock Using parallel computation to improve independent
  {M}etropolis--{H}astings based estimation.
\newblock {\em J. Comput. Graph. Statist.}, 20(3):616--635.

\bibitem[Jacob and Thiery, 2015]{jacob:thiery:2015}
Jacob, P.~E. and Thiery, A.~H. (2015).
\newblock On nonnegative unbiased estimators.
\newblock {\em Ann. Statist.}, 43(2):769--784.

\bibitem[Johansen et~al., 2012]{johansen:whiteley:doucet:2012}
Johansen, A.~M., Whiteley, N., and Doucet, A. (2012).
\newblock Exact approximation of {R}ao--{B}lackwellised particle filters.
\newblock {\em IFAC Proceedings Volumes}, 45(16):488 -- 493.
\newblock 16th IFAC Symposium on System Identification.

\bibitem[Kong et~al., 2003]{kong:mccullagh:meng:2003}
Kong, A., McCullagh, P., Meng, X.~L., Nicolae, D., and Tan, Z. (2003).
\newblock A theory of statistical models for {M}onte-{C}arlo integration.
\newblock {\em J. Royal Statist. Society Series B}, 65(3):585--618.

\bibitem[Liao, 1998]{liao:1998}
Liao, J. (1998).
\newblock Variance reduction in {G}ibbs sampler using quasi-random numbers.
\newblock {\em J. Comput. Graph. Statist.}, 7(3):253--266.

\bibitem[Lindsten, 2011]{Lindsten:2011}
Lindsten, F. (2011).
\newblock {\em Rao-{B}lackwellised particle methods for inference and
  identification}.
\newblock PhD thesis, University of Link{\"o}ping, Sweden.

\bibitem[Lindsten et~al., 2011]{lindsten:schon:olsson:2011}
Lindsten, F., Sch{\"o}n, T.~B., and Olsson, J. (2011).
\newblock An explicit variance reduction expression for the
  {R}ao--{B}lackwellised particle filter.
\newblock {\em IFAC Proceedings Volumes}, 44(1):11979 -- 11984.
\newblock 18th IFAC World Congress.

\bibitem[Liu and Chen, 1998]{liu:chen:1998}
Liu, J. and Chen, R. (1998).
\newblock Sequential {M}onte-{C}arlo methods for dynamic systems.
\newblock {\em J. American Statist. Assoc.}, 93:1032--1044.

\bibitem[Liu et~al., 1994]{liu:wong:kong:1994}
Liu, J., Wong, W., and Kong, A. (1994).
\newblock Covariance structure of the {G}ibbs sampler with applications to the
  comparisons of estimators and sampling schemes.
\newblock {\em Biometrika}, 81:27--40.

\bibitem[Malefaki and Iliopoulos, 2008]{malefaki:iliopoulos:2008}
Malefaki, S. and Iliopoulos, G. (2008).
\newblock On convergence of importance sampling and other properly weighted
  samples to the target distribution.
\newblock {\em J. Statist. Plann. Inference}, 138:1210--1225.

\bibitem[Marin and Robert, 2010]{marin:robert:2010b}
Marin, J. and Robert, C. (2010).
\newblock On resolving the {S}avage--{D}ickey paradox.
\newblock {\em Electron. J. Statist.}, 4:643--654.

\bibitem[Marin and Robert, 2011]{marin:robert:2010}
Marin, J. and Robert, C. (2011).
\newblock Importance sampling methods for {B}ayesian discrimination between
  embedded models.
\newblock In Chen, M.-H., Dey, D., M{\"u}ller, P., Sun, D., and Ye, K.,
  editors, {\em Frontiers of Statistical Decision Making and {B}ayesian
  Analysis}, pages 513--527. Springer-Verlag, New York.

\bibitem[Marinari and Parisi, 1992]{marinarietparisi92}
Marinari, E. and Parisi, G. (1992).
\newblock Simulated tempering: A new {M}onte {C}arlo schemes.
\newblock {\em Europhysics letters}, 19:451--458.

\bibitem[McKeague and Wefelmeyer, 2000]{mckeague:wefelmeyer:2000}
McKeague, I. and Wefelmeyer, W. (2000).
\newblock {M}arkov chain {M}onte {C}arlo and {R}ao--{B}lackwellisation.
\newblock {\em J. Statist. Plann. Inference}, 85:171--182.

\bibitem[Metropolis et~al., 1953]{metropolis:1953}
Metropolis, N., Rosenbluth, A.~W., Rosenbluth, M.~N., Teller, A.~H., and
  Teller, E. (1953).
\newblock Equations of state calculations by fast computing machines.
\newblock {\em J. Chem. Phys.}, 21:1087--1092.

\bibitem[Mira et~al., 2001]{mira:moeller:roberts:2001}
Mira, A., M{{\o}}ller, J., and Roberts, G. (2001).
\newblock Perfect slice samplers.
\newblock {\em J. Royal Statist. Society Series B}, 63:583--606.

\bibitem[Neal, 1999]{neal:1999}
Neal, R. (1999).
\newblock Erroneous results in ``{M}arginal likelihood from the {G}ibbs
  output``.
\newblock Technical report, University of Toronto.

\bibitem[Owen and Zhou, 2000]{owen:zhou:00}
Owen, A. and Zhou, Y. (2000).
\newblock Safe and effective importance sampling.
\newblock {\em J. American Statistical Association}, 95:135--143.

\bibitem[Pearl, 1987]{pearl:1987}
Pearl, J. (1987).
\newblock Evidential reasoning using stochastic simulation in causal models.
\newblock {\em Artificial Intelligence}, 32:247--257.

\bibitem[Perron, 1999]{perron:1999}
Perron, F. (1999).
\newblock Beyond accept--reject sampling.
\newblock {\em Biometrika}, 86(4):803--813.

\bibitem[Rao, 1945]{rao:1945}
Rao, C. (1945).
\newblock Information and the accuracy attainable in the estimation of
  statistical parameters.
\newblock {\em Bulletin of the Calcutta Mathematical Society}, 37:81--91.

\bibitem[Robert and Casella, 2004]{robert:casella:2004}
Robert, C. and Casella, G. (2004).
\newblock {\em {M}onte {C}arlo Statistical Methods}.
\newblock Springer-Verlag, New York, second edition.

\bibitem[Robert and Marin, 2008]{robert:marin:2008}
Robert, C. and Marin, J.-M. (2008).
\newblock On some difficulties with a posterior probability approximation
  technique.
\newblock {\em Bayesian Analysis}, 3(2):427--442.

\bibitem[Robert and Wraith, 2009]{robert:wraith:2009}
Robert, C. and Wraith, D. (2009).
\newblock Computational methods for {B}ayesian model choice.
\newblock In Goggans, P.~M. and Chan, C.-Y., editors, {\em MaxEnt 2009
  proceedings}, volume 1193. AIP.

\bibitem[Roberts and Rosenthal, 1999]{roberts:rosenthal:1999}
Roberts, G. and Rosenthal, J. (1999).
\newblock Convergence of slice sampler {M}arkov chains.
\newblock {\em J. Royal Statist. Society Series B}, 61:643--660.

\bibitem[Sahu and Zhigljavsky, 1998]{sahu:zhigljavsky:1998}
Sahu, S. and Zhigljavsky, A. (1998).
\newblock Adaptation for self regenerative {MCMC}.
\newblock Technical report, Univ. of Wales, Cardiff.

\bibitem[Sahu and Zhigljavsky, 2003]{sahu:zhigljavsky:2003}
Sahu, S. and Zhigljavsky, A. (2003).
\newblock Self regenerative {M}arkov chain {M}onte {C}arlo with adaptation.
\newblock {\em Bernoulli}, 9:395--422.

\bibitem[Spiegelhalter et~al., 1995]{spiegelhalter:thomas:best:gilks:1995}
Spiegelhalter, D., Thomas, A., Best, N., and Gilks, W. (1995).
\newblock {BUGS}: {B}ayesian inference using {G}ibbs sampling.
\newblock Technical report, Medical Research Council Biostatistics Unit,
  Institute of Public Health, Cambridge University.

\bibitem[Tanner and Wong, 1987]{tanner:wong:1987}
Tanner, M. and Wong, W. (1987).
\newblock The calculation of posterior distributions by data augmentation.
\newblock {\em J. American Statist. Assoc.}, 82:528--550.

\bibitem[Vehtari et~al., 2019]{vehtari:simpson:etal:2019}
Vehtari, A., Simpson, D., Gelman, A., Yao, Y., and Gabry, J. (2019).
\newblock Pareto smoothed importance sampling.
\newblock arXiv:1507.02646.

\end{thebibliography}

\end{document}